\documentclass[amsmath,amssymb,superscriptaddress]{revtex4}
\usepackage{graphicx}% Include figure files
\usepackage{dcolumn}% Align table columns on decimal point
\usepackage{bm}% bold math
\usepackage[colorlinks=true, citecolor=blue, urlcolor = blue, linkcolor= red, bookmarks=true]{hyperref}
\begin{document}

\title{Nonsingular black hole chemistry}

\author{Arun Kumar}
\email{arunbidhan@gmail.com}
\affiliation{Centre for Theoretical Physics,
Jamia Millia Islamia, New Delhi 110025,
India}
\author{Sushant G. Ghosh}
\email{sgghosh@gmail.com}
\affiliation{Centre for Theoretical Physics,
Jamia Millia Islamia, New Delhi 110025,
India} 
\affiliation{Astrophysics and Cosmology Research Unit,
School of Mathematics, Statistics and Computer Science,
University of KwaZulu-Natal, Private Bag X54001,
Durban 4000, South Africa}
\author{Sunil D. Maharaj}
\email{maharaj@ukzn.ac.za}
\affiliation{Astrophysics and Cosmology Research Unit,
School of Mathematics, Statistics and Computer Science,
University of KwaZulu-Natal, Private Bag X54001,
Durban 4000, South Africa}

\begin{abstract}
We study the nonsingular black hole in Anti de-Sitter background taking the negative cosmological constant  as the pressure of the system. We investigate the horizon structure, and find the critical values $m_0$ and $\tilde{k}_0$, such that $m>m_0$ (or $\tilde{k}<\tilde{k}_0$) corresponds to a black solution with two horizons, namely the Cauchy horizon $x_-$ and the event horizon $x_+$. For $m=m_0$ (or $\tilde{k}=\tilde{k}_0$), there exist an extremal black hole with degenerate horizon $x_0=x_{\pm}$ and for $m<m_0$ (or $\tilde{k}>\tilde{k}_0$), no black hole solution exists. In turn, we calculate the thermodynamical properties and by observing the behaviour of Gibb's free energy and specific heat, we find that this black hole solution exhibits first order (small to large black hole) and second order phase transition. Further, we study the $P-V$ criticality of system and then calculate the critical exponents showing that they are the same as those of the Van der Waals fluid.

\end{abstract}

\pacs{04.20.Jb, 04.70.Bw, 04.40.Nr}% PACS, the Physics and Astronomy
% Classification Scheme.
\keywords{Regular black holes, thermodynamics, P-V criticality }%Use showkeys class option if keyword
%display desired
\maketitle

\section{Introduction} 

 The theoretical prediction of black holes by Einstein's general relativity opened a new door for research on very fascinating astrophysical objects, which are ultimate sponges of nature, absorbing all matter and emitting nothing. Bekenstein and Hawking made a breakthrough on the thermodynamical aspect of black holes, by relating the temperature and entropy of black holes at the event horizon, respectively, with the surface gravity at the event horizon and the area of the event horizon \cite{1}. Following this pioneering work, the thermodynamical properties of black holes have been studied extensively \cite{2}. Hawking and Page \cite{3} in 1983, discovered that the black hole in anti de Sitter (AdS) spacetimes, undergoes a phase transition between pure radiation and a stable black hole, known as the Hawking-Page phase transition. That discovery by Hawking and Page led the research society to show much interest in the study of black holes in AdS spacetimes, which continues today.
 
 Recently, a very interesting proposal \cite{KuM14}, added a new chemical aspect to the thermodynamical study of black holes, by interpreting the negative cosmological constant $\Lambda$ of AdS spacetime as the positive pressure of the thermodynamical system, via $P=-\Lambda/8\pi$ and with it the concept of black hole volume also gets introduced. This extension leads to the modification in the first law of black hole thermodynamics with the inclusion of the term '$VdP$' \cite{dk,5} and as a result the black hole mass is interpreted as enthalpy instead of the internal energy. Kubiznak and Mann \cite{KuM14} studied the '$P-V$' criticality of charged black holes in AdS spacetime, and found that the critical exponents coincide with those of Van der Waals fluid \cite{dr1}. Following that treatment $P-V$ criticality for various black holes has been studied \cite{dh,7}. This new emerging subfield of black hole thermodynamics was hence termed as black hole chemistry \cite{dr,8}.
 
 The existence of singularities is one of the most discussed areas in general relativity. There has been a lot of efforts to overcome the problems with singularities since the time of the introduction of general relativity by Einstein. Bardeen \cite{9} was the first to propose a nonsingular model of black holes by coupling general relativity minimally with nonlinear electrodynamics (NED). Since then, many black hole models with regular core have been introduced \cite{ak,10}. In this paper, our goal is to study the chemistry of a nonsingular black hole \cite{sg}, which is parameterized by the mass $M$ and a parameter $k$. The nonsingular black hole line element reads
 \begin{equation}
 ds^2=-\left(1-\frac{2Me^{\frac{-k}{r}}}{r}\right)dt^2+\frac{1}{\left(1-\frac{2Me^{\frac{-k}{r}}}{r}\right)}dr^2+r^2(d\theta^2+{\sin}^2\theta d\phi^2),
 \end{equation}
 where the parameter $k$ is related to the NED charge $e$, via, $e^2=2Mk$ . This article is organized as follows. Sec. 2 is devoted to the nonsingular black holes solution in AdS spacetime, followed by discussion of curvature scalars and horizon structure. In Sec. 3, we derive expressions for all the thermodynamical quantities, like temperature $T_+$, entropy $S_+$, Gibb's free energy $G_+$ and the specific heat $C_+$. We also derive the equation of state and discuss the $P-V$ criticality. We calculate the values of critical exponents $\alpha, \beta, \gamma $ and $\delta$ in Sec. 4. Finally, we conclude the paper in Sec. 5. In this paper, we use units in which $G = \hslash  = c = 1$ .

\section{Nonsingular-anti-deSitter black hole}
\label{BAdS_section}

 We first generalize the nonsingular black holes \cite{sg} to anti-de Sitter spacetime. The solution we are interested in, can be derived from general relativity minimally coupled to nonlinear electrodynamics with negative $\Lambda$-term given by the action \cite{cai}:
\begin{equation}\label{action}
\mathcal{S}= \frac{1}{16\pi} \int \mathrm{d} ^4x \sqrt{-g} \left( R+6l^{-2}-\mathcal{L}(F)\right) \,,
\end{equation}
where $R$ is the Ricci scalar, $g$ is the determinant of the metric tensor, $l$ is the positive AdS radius and $\mathcal{L}(F)$ is the Lagrangian density of nonlinear electrodynamics which is a function of $F=1/4 F^{\mu\nu}F_{\mu\nu}$ with $F_{\mu\nu}=\partial_{\mu}A_{\nu}-\partial_{\nu}A_{\mu}$, the strength tensor of nonlinear electrodynamics. By varying the action (\ref{action}), we derive the following equations of motion
\begin{eqnarray} \label{field_eq1}
G_{\mu\nu} -3l^{-2} g_{\mu\nu} &=& T_{\mu\nu} \equiv  2 \left( \frac{\partial \mathcal{L}(F)}{\partial F} F_{\mu \lambda} F _{\nu}{}^{\lambda} - g_{\mu\nu} \mathcal{L}(F) \right) \,, \\ 
 &&\nabla_{\mu} \left( \frac{\partial \mathcal{L}(F)}{\partial F} F^{\mu\nu}  \right)=0 \,, \label{field_eq2}
\end{eqnarray}
where $G_{\mu\nu}$ is the Einstein tensor. The solution that we are interested in can be obtained from the Lagrangian density
\begin{equation} \label{L_term}
\mathcal{L}(F) = Fe^{\frac{-k}{e}(2e^2F)^{\frac{1}{4}}} ,
\end{equation}
where $e$ is a magnetic charge and $k>0$ is a constant. We consider the Maxwell field tensor \cite{sg1} 
\begin{equation}
F_{\mu\nu}=2\delta^{\theta}_{[\mu}\delta^{\phi}_{\nu]}e(r)\sin\theta.
\end{equation} By using Eq. (\ref{field_eq2}), we get the condition $e'(r)\sin\theta d r\wedge d\theta \wedge d\phi=0$, which implies $e(r)=\text{constant}=e$, and we get.
\begin{equation} \label{field}
F_{\theta\phi}=e\sin\theta \qquad\qquad \text{and} \qquad\qquad F=\frac{e^2}{2r^4}.
\end{equation}
Now, by inserting Eq. (\ref{field}) in Eq. (\ref{L_term}), we obtain the Lagrangian density
\begin{equation}
\mathcal{L}(F)=\mathcal{L}(r)\equiv\frac{e^2}{2r^4}e^{-k/r}.
\end{equation} To obtain the metric of the nonsingular black hole in AdS spacetime, we assume the static and spherically symmetric line element of the form \cite{ghosh}
\begin{equation}\label{metric}
ds^2=-f(r)dt^2+f(r)^{-1}dr^2+r^2d\Omega^2.
\end{equation} Here $f(r)$ is the metric function to be determined and $d\Omega^2=d\theta^2+\sin^2\theta d\phi^2$.
Using the line element (\ref{metric}), the $(t,t)$ component of the Einstein field equation reads
\begin{equation}\label{eqn}
\frac{1}{r^2}[rf'+f-1]-\frac{3}{l^2}=\frac{e^2}{r^4}e^{-k/r},
\end{equation} where a prime denotes the derivative with respect to $r$. Eq (\ref{eqn}) can be easily integrated to give 

\begin{equation} \label{mf}
f(r) =1 -\frac{2 M e^{-k/r}}{r}+\frac{r^2}{l^2},
\end{equation}
where the mass of black hole $M$ and magnetic charge $e$ are related by parameter $k$, via $e^2=2Mk.$ Thus the metric for nonsingular black holes in AdS spacetime reads
\begin{equation}\label{mf2}
ds^2=-\left(1 -\frac{2 M e^{-k/r}}{r}+\frac{r^2}{l^2}\right)dt^2+\frac{1}{\left(1 -\frac{2 M e^{-k/r}}{r}+\frac{r^2}{l^2}\right)}dr^2+r^2d\Omega^2.
\end{equation}
 The solution (\ref{mf}) encompasses the Schwarzschild-AdS \cite{KuM17,ak} as a special case in the absence of nonlinear electrodynamics ($k=0$). Also, the nonsingular black hole (\ref{mf}) without cosmological term ($1/l^2=0$) has been extensively studied in \cite{sg}. 

\begin{figure}[h!h!] 
\begin{tabular}{c c c c}
\includegraphics[width=0.5 \textwidth]{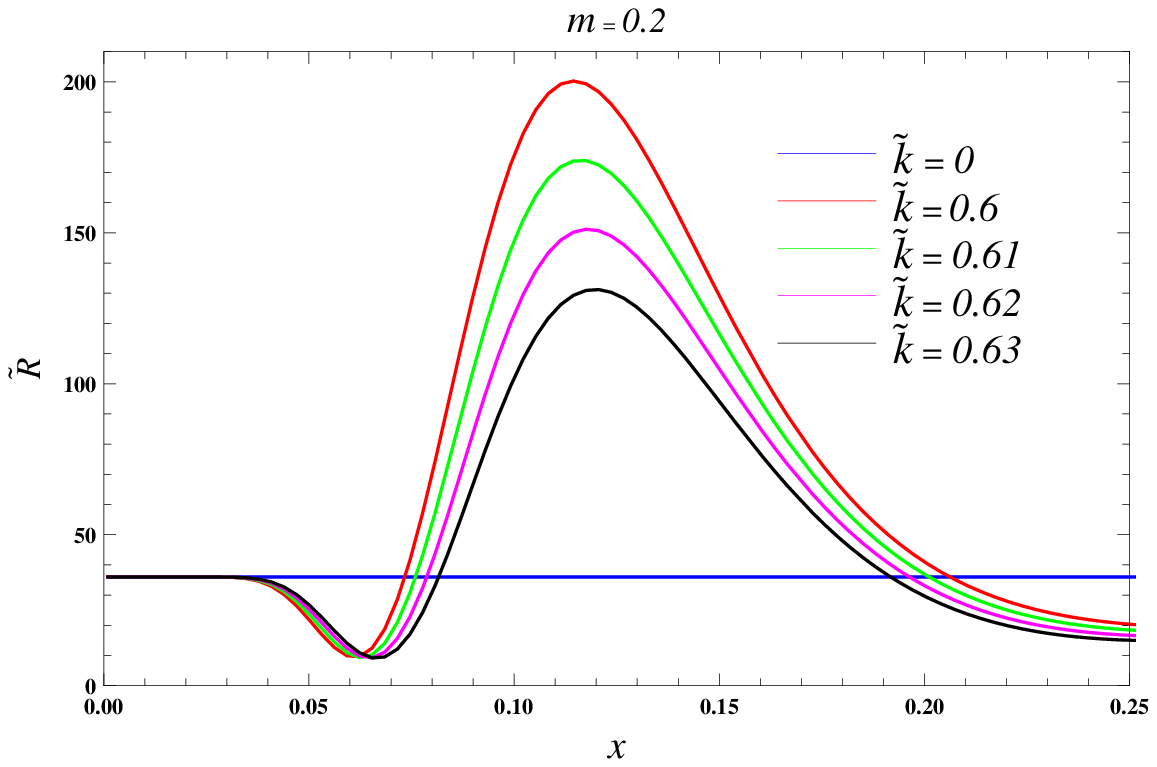}
\includegraphics[width=0.5 \textwidth]{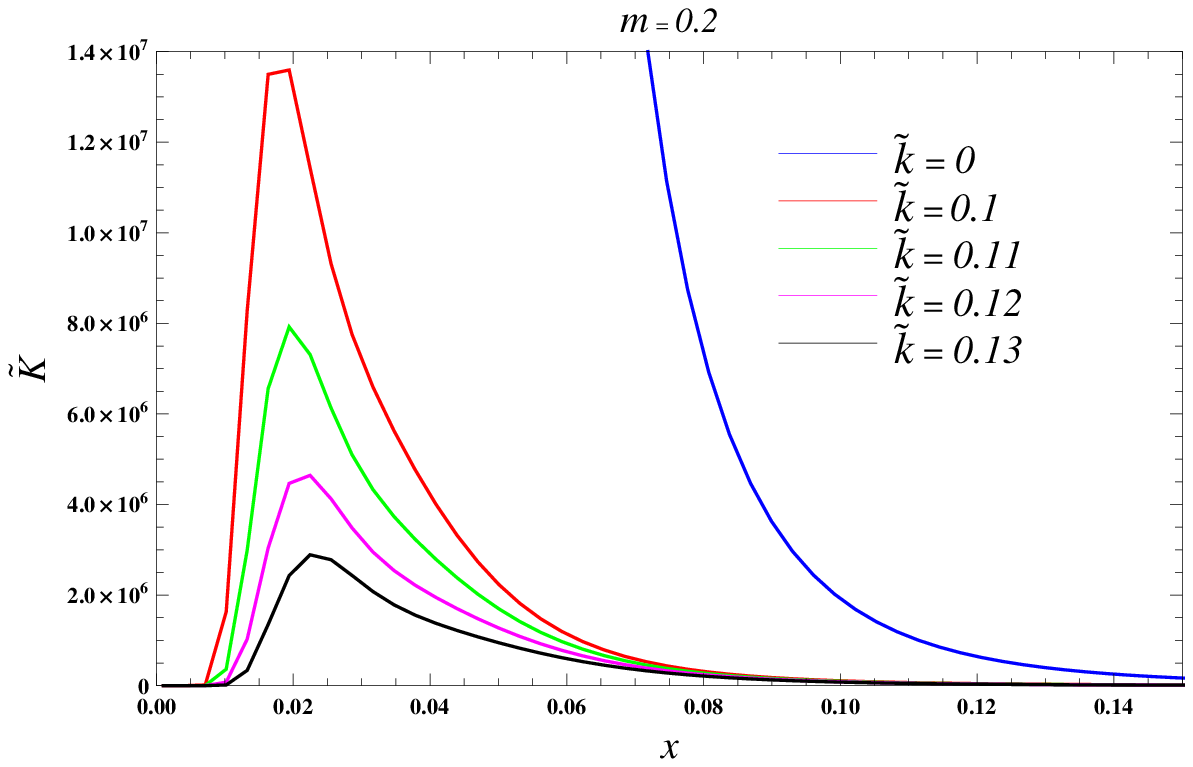}
\end{tabular}
\caption{ The plot of Ricci square ($\mathcal{\tilde{R}}$) and Kretshmann scalar ($\mathcal{\tilde{K}}$) with $x$ for different values of deviation parameter $\tilde{k}$.}
\label{scalar}
\end{figure}

To consider the chemistry of nonsingular black holes, we rewrite solution (\ref{mf}) in $l$ units 
\begin{equation} \label{mf1}
f(x) =  1 -\frac{2 m e^{-\tilde{k}/x}}{x}+{x^2},
\end{equation}
where
\begin{equation} \label{dim_var}
x=r/l\,, \ \ \ \ m=M/l \,, \ \ \ \  \tilde{k}=k/l \,.
\end{equation} 
Asymptotically ($x>>\tilde{k}$), the solution \eqref{mf1} mimics the well known Reissner-Nördstrom-AdS black hole
\begin{equation}
f(x)\approx 1-\frac{2m}{x}-\frac{e^2}{x^2}+x^2,
\end{equation} where $e$ is the charge.
While for ($x \to 0$), metric function \eqref{mf1} reduces to
\begin{equation}
f(x)\approx 1+x^2,
\end{equation} which signifies that the black hole solution near the origin becomes the AdS vacuum solution. To examine the regularity of the solution (\ref{mf}), it is necessary to study the behaviour of scalar invariants, Ricci square ($\mathcal{R} = R_{ab}R^{ab}$) and Kretschmann scalar ($\mathcal{K} = R_{abcd}R^{abcd}$):
\begin{eqnarray}
\mathcal{\tilde{R}}&=&\mathcal{R}l^4=36-\frac{2m\tilde{k}^2e^{-\frac{\tilde{k}}{x}}}{x^{10}}\left[6x^5-me^{-\frac{\tilde{k}}{x}}(8x^2-4\tilde{k}x+\tilde{k}^2)\right],\nonumber\\
\mathcal{\tilde{K}}&=&\mathcal{K}l^4=24-\frac{4me^{-\frac{\tilde{k}}{x}}}{x^{10}}\left[2\tilde{k}^2x^5-me^{-\frac{\tilde{k}}{x}}\left(12x^4-24\tilde{k}x^2(x-\tilde{k})-\tilde{k}^3(8x-\tilde{k})\right)\right].
\end{eqnarray}
 For a nonsingular-AdS black hole these invariants are well behaved everywhere including $x=0$ (cf. Fig. \ref{scalar}), signifying the regularity of the solution (\ref{mf}).

\begin{figure}[h!h!] 
\begin{tabular}{c c c c}
\includegraphics[width=0.5 \textwidth]{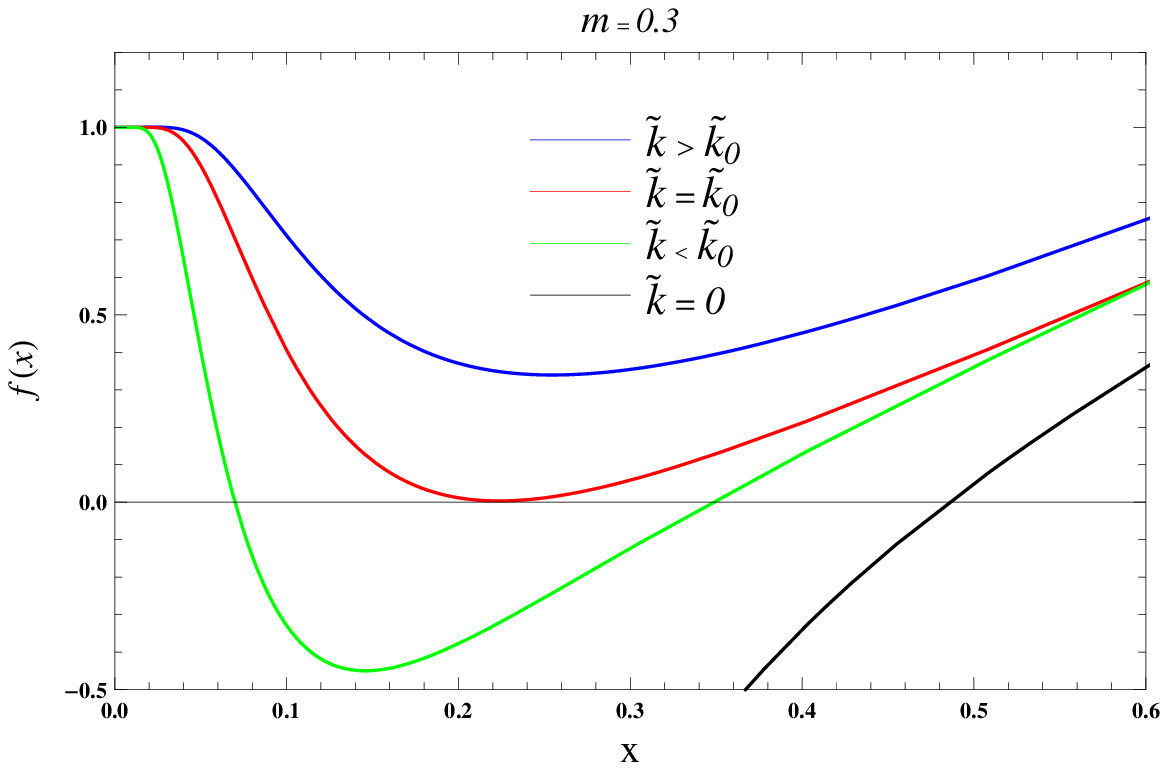}
\includegraphics[width=0.5 \textwidth]{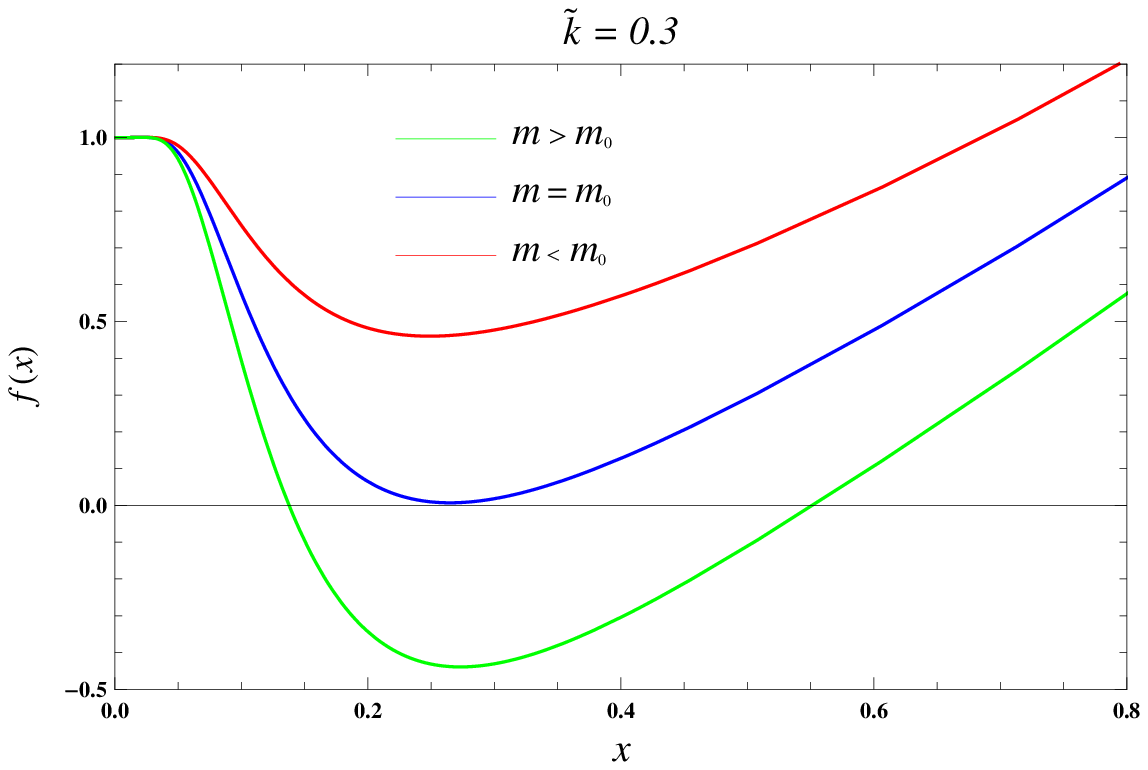}
\end{tabular}
\caption{ The metric function $f(x)$ for one-horizon and double-horizon nonsingular black holes configurations in Anti de Sitter space.}
\label{fig:Bpotential}

\end{figure}

The metric has a coordinate singularity at $f(r)=0$, implying the existence of a horizon. It turns out that, for a given value of $\tilde{k}$ (or $m$), there exists a critical value $m_0$ (or $\tilde{k_0}$), such that for $m>m_0$ (or $\tilde{k}<\tilde{k_0}$), $f(x)=0$ has two simple zeros, and no zero for $m<m_0$ (or $\tilde{k}>\tilde{k_0}$) (cf. Fig.~\ref{fig:Bpotential} ). The case $m=m_0$ (or $\tilde{k}=\tilde{k_0}$) corresponds to an extremal black hole with degenerate horizon ($x_+=x_-$), with $x_+$ and $x_-$, respectively, mean event and Cauchy horizons.
 \begin{figure}[h!h!] 
\begin{tabular}{c c c c}
\includegraphics[width=0.6 \textwidth]{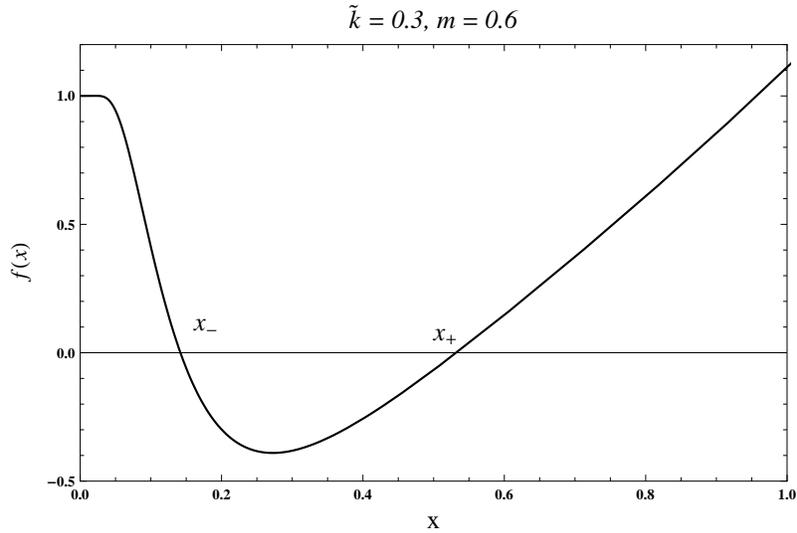}
\end{tabular}
\caption{ Metric function $f(x)$ for nonsingular-AdS black hole.}
\label{f}

\end{figure}

To find $x_0$ and $m_0$, we note that the extremal black holes are defined by 
\begin{equation}\label{eq1}
f(x)=0=\frac{\partial f(x)}{\partial x}|_{x=x_0} \,,
\end{equation}
Solving Eq. \eqref{eq1}, we obtain
\begin{eqnarray}\label{r0}
x_0 = \frac{1}{9\beta_1} \left[\tilde{k}^2+\beta_1 \tilde{k}+\beta_1^2 -9\right]~~~\text{with}~~~\beta_1=\left[\tilde{k}^3+108\tilde{k}+9\sqrt{3\tilde{k}^4+141\tilde{k}^2+9}\right]^{\frac{1}{3}},
\end{eqnarray} and
\begin{equation} \label{B_mass}
m_0=  e^{9\beta_1\tilde{k}/\tilde{k}^2+\beta_1 \tilde{k}+\beta_1^2 -9} \left[\frac{\left(81\beta_1^2+(\tilde{k}^2+\beta_1 \tilde{k}+\beta_1^2 -9)^2\right)(\tilde{k}^2+\beta_1 \tilde{k}+\beta_1^2 -9)}{2(9\beta_1)^3}\right].\nonumber
\end{equation} 
 For our example in Fig.~\ref{fig:Bpotential}, for $\tilde{k}=0.3\,$, the black hole exists if $m \geq m_0 \approx 0.443$, which provides a lower bound to the horizon radius of the black hole $x \geq x_0 \approx 0.26\,$.
\begin{figure}[h!h!] 
\begin{tabular}{c c c c}
\includegraphics[width=0.5 \textwidth]{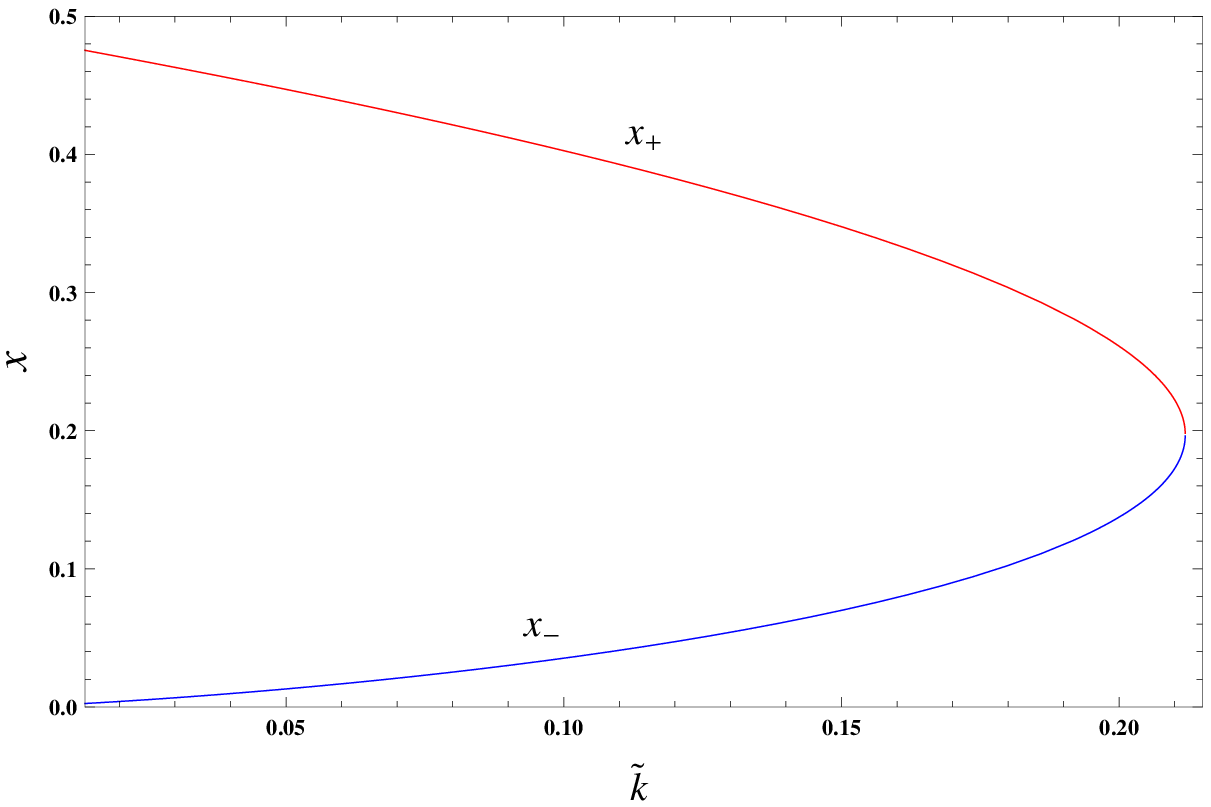}
\includegraphics[width=0.5 \textwidth]{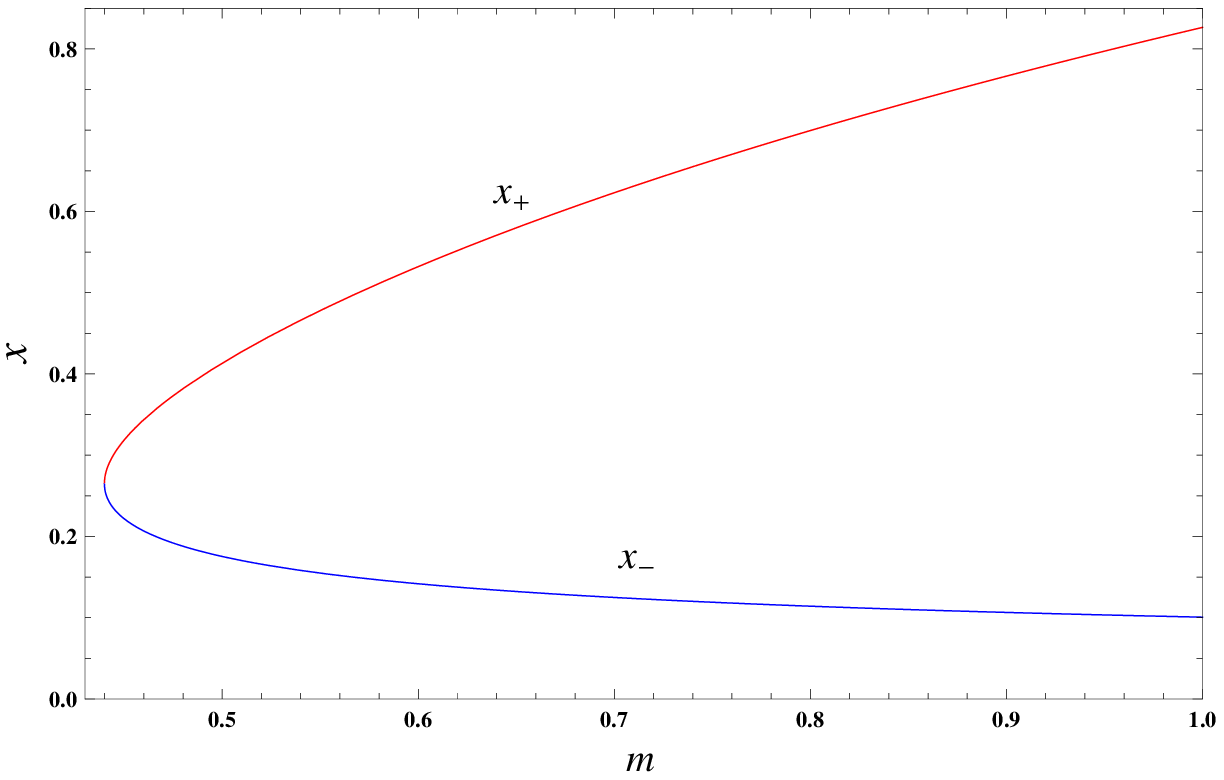}
\end{tabular}
\caption{ Horizons for $m=0.3$ (left) and for $\tilde{k}=0.3$ (right).}
\label{fig:Bpotential1}
\end{figure}

\section{ nonsingular black hole chemistry}
\label{BAdS_chemistry}
We shall analyze the thermodynamical quantities associated with the black hole horizon $x_+$. We can extend the known thermodynamic relations to the case in which a specific microscopic structure of the quantum spacetime is prescribed \cite{NiT11,Nic10,ak1}. From the solution of equation $f(r_+)=0$, where $r_+$ is the horizon radius, we obtain the black hole mass as 
\begin{equation} \label{B_mass}
M= \frac{r_+}{2} \left[e^{k/r}(1+\frac{r_+^2}{l^2})\right],
\end{equation}
which in terms of $x_+$, becomes
\begin{equation} \label{B_mass1}
m= \frac{x_+}{2} \left[e^{\tilde{k}/x_+}(1+x_+^2)\right].
\end{equation}
 The Hawking temperature, associated with the black hole horizon can be obtained as in \cite{hk1,hk2}, which reads
\begin{equation} \label{B_temp0}
T_+=\frac{1}{4\pi} \frac{ \partial f(r)}{\partial r}|_{r=r_+} = \frac{1}{4\pi r_+}\left[1+\frac{3r^2_+}{l^2}-\frac{k}{r_+}(1+\frac{r^2_+}{l^2})\right], \,
\end{equation}
by using Eq. \eqref{dim_var}. Rewriting the black hole temperature in terms of dimensionless quantities, we obtain
\begin{equation} \label{B_temp}
\tilde{T} = \frac{1}{4\pi  x_+}\left[1+3x^2_+-\frac{\tilde{k}}{x_+}(1+x^2_+) \right].
\end{equation}
The temperature $\tilde{T}>0$, if $3x^3_+-\tilde{k}x^2_+ + x_+-\tilde{k} \geq 0$ or $x_+ \geq \frac{1}{9\beta} \left[\tilde{k}^2+\beta \tilde{k}+\beta^2 -9\right] = x_0\,$.
\begin{figure}[h!h!] 
\begin{tabular}{c c c c}
\includegraphics[width=0.5 \textwidth]{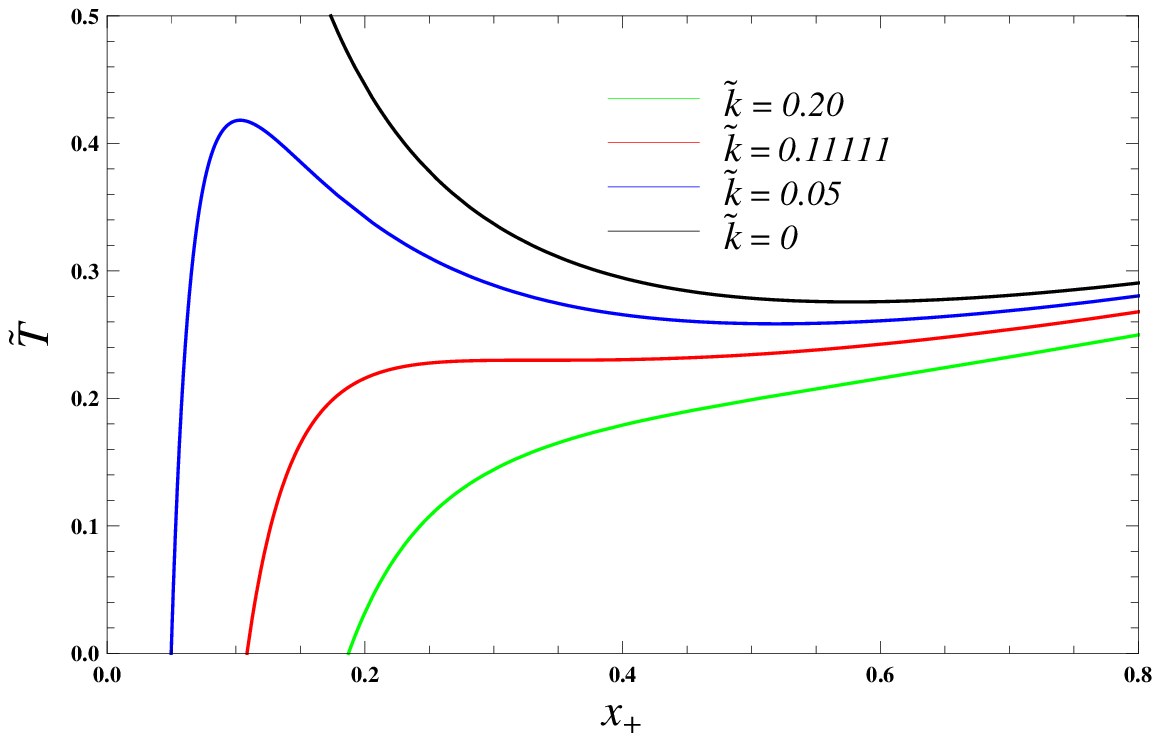}
\includegraphics[width=0.5 \textwidth]{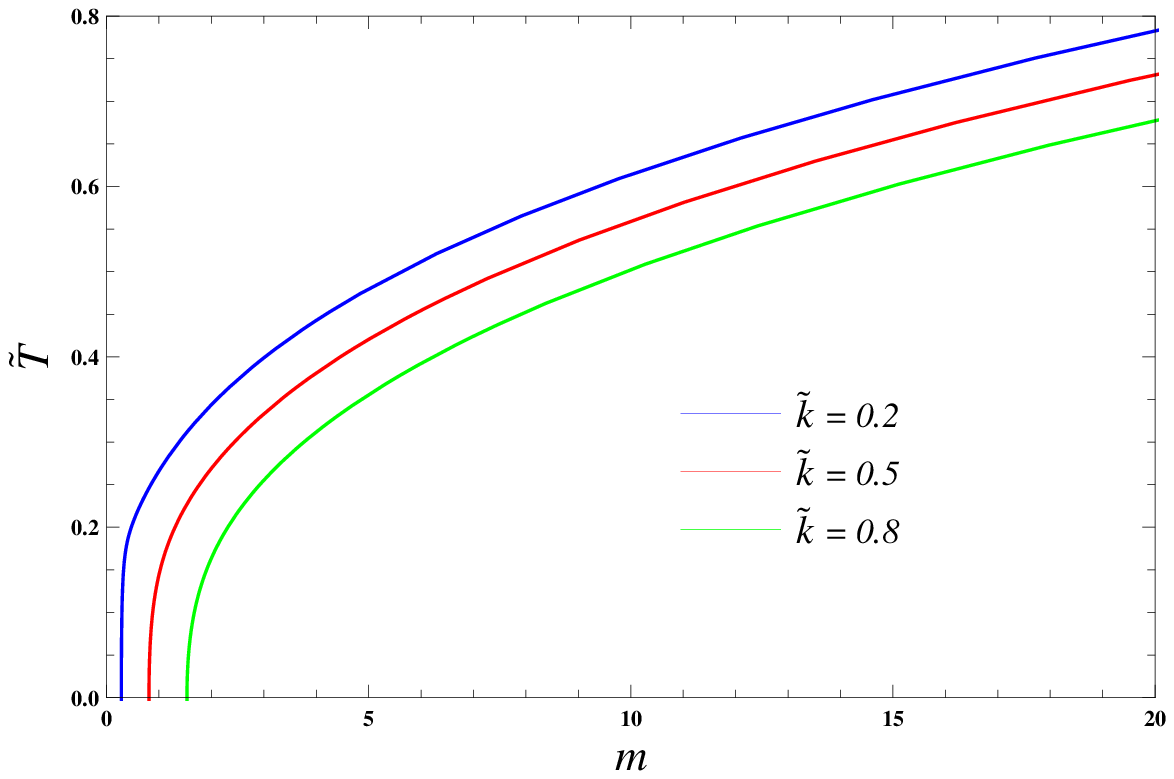}
\end{tabular}
\caption{ The temperature $\tilde{T}$ \textit{vs} horizon $x_+$ (left) and mass $m$ (right).}
\label{fig:Btemp}
\end{figure}
The position of local extrema of the Hawking temperature (\ref{B_temp}) can be identified as the roots of $\partial \tilde{T}/ \partial x |_{x_+=x_e}=0$ or as the zeros of the polynomial $q(x)=3 x^3-x+2 k=0$. The numerical analysis of solution of $q(x)=0$, leads us to find out a critical value of $\tilde{k_c}=0.11111$, such that for $\tilde{k}=0.05<\tilde{k_c}$, we have local maxima of temperature at $x_+=0.10330$ and minima at $x_+=0.51872$. For $\tilde{k}=0.20 >\tilde{k_c}$, the Hawking temperature increases monotonically. It is interesting to note that when $\tilde{k}=\tilde{k_c}$, both the extremal converges at $x_+=0.33247$ (cf. Fig.~\ref{fig:Btemp}).
To find the entropy of the black hole associated with horizon, we use the first law of thermodynamics \cite{ghosh}
\begin{equation}\label{th1}
dM_+=T_+ dS_+.
\end{equation}
Substituting Eqs. \eqref{B_mass1} and \eqref{B_temp} in Eq. \eqref{th1}, and integrating, the entropy of the nonsingular black hole takes the form
\begin{equation} \label{B_s}
\tilde{S}=\frac{S}{l^2}=\int \limits^{x_+}_{0} 2 \pi  x e^{\tilde{k}/x} dx= \frac{A}{4}\left[e^{\tilde{k}/x_+} \left(1+\tilde{k}{x_+}\right)-\frac{\tilde{k}^2}{x_+^2} \text{Ei}\left[\frac{\tilde{k}}{x_+}\right]\right],  
\end{equation}
where Ei is an exponential integral function and $A=4\pi x_+^2$. The standard area law $S=A/4$ is modified due to nonlinear electrodynamics.
Because of the additional parameter $\tilde{k}$ in nonsingular black holes, the first law of black hole thermodynamics modifies to
\begin{equation}\label{th2}
dM=TdS + \bar{\phi}dk,
\end{equation} where $\bar{\phi}$ is the conjugate of the charge parameter $k$. The first law of black hole thermodynamics is analogous to the first law of standard thermodynamics,
\begin{equation}\label{th3}
dE=TdS-PdV+ \text{work terms},
\end{equation} except for the absence of the pressure-volume term in (\ref{th2}). This problem of the absence of the $P-V$ term has been addressed by interpreting the mass of the black hole as the enthalpy of spacetime. The consideration of the Smarr relation \cite{mm,dk} gave birth to this idea, which has the following form for static asymptotically flat black hole  \cite{smr}
\begin{equation}
M=2TS'.
\end{equation}
The inclusion of the cosmological constant ($\Lambda$) and the charge parameter ($k$), necessarily modifies the Smarr relation as can be shown from geometric arguments \cite{bp,dk}. So, now considering $M=M(S',\Lambda,k)$, Euler's theorem implies 
\begin{equation}\label{smarr}
M=2 \frac{\partial M}{\partial S'}S'-2 \frac{\partial M}{\partial \Lambda }\Lambda+\bar{\phi}k,
\end{equation} with $\partial M/\partial S'=T$, and $P=-\Lambda/8\pi$ has been considered as a thermodynamical variable \cite{jc,tp}, with it's conjugate $V=\partial M/\partial P$. Using these, we get the general form of the Smarr formula \cite{dr,jz}
\begin{equation}\label{smarr2}
M=2TS'-2PV+\bar{\phi}k.
\end{equation} The cosmological prospective, that the negative cosmological constant induces a vacuum pressure, leads to the identification of the quantity $P$ as thermodynamical pressure. Hence, the modified first law of black hole thermodynamics can be written as
\begin{equation}\label{th3}
dM=TdS+VdP+\bar{\phi}dk.
\end{equation} The thermodynamical quantities which satisfy the new first law of thermodynamics (\ref{th3}) for nonsingular black hole are

\begin{eqnarray}\label{new_th}
M &=&  \frac{r_+}{2} \left[e^{k/r}(1+\frac{r_+^2}{l^2})\right] \,, \ \ \  \bar{\phi} = \frac{ e^{k/r_+}}{2 } \left[1+\frac{r_+^2}{l^2}\right]\nonumber,
\\ T &=& \frac{1}{4\pi r_+}\left[1+\frac{3r^2_+}{l^2}-\frac{k}{r_+}(1+\frac{r^2_+}{l^2})\right]  \,, \ \ \ V = \frac{4}{3} \pi  r_+^3 e^{k/r_+} \,,  \  \ \ P = \frac{3}{8\pi l^2}\nonumber, \\   S &=& \pi r_+^2 \left[ e^{k/r_+} (1+\frac{k}{r_+})-\frac{k^2}{r_+^2} \text{Ei}\left[\frac{k}{r_+}\right]\right]  .
\end{eqnarray}
 When we take the limit ($k=0$), all the thermodynamic quantities given in Eq. (\ref{new_th}), reduce to the corresponding thermodynamical quantities of Schwarzschild-AdS black holes \cite{KuM17},
 \begin{eqnarray}\label{new_th1}
M &=&  \frac{r_+}{2} \left[1+\frac{r_+^2}{l^2}\right],\,\, \,\,\,  \bar{\phi} = \frac{ 1}{2 } \left[1+\frac{r_+^2}{l^2}\right],\,\,\,\,\,
T = \frac{1}{4\pi r_+}\left[1+\frac{3r^2_+}{l^2}\right],\nonumber\\ \ \ \ V&=& \frac{4}{3} \pi  r_+^3 \,,  \  \ \ P = \frac{3}{8\pi l^2},\,\,\,   S = \pi r_+^2.
\end{eqnarray}

Following \cite{MiX17,Lia17}, the parameters $k$ and $\bar{\phi}\,$ can be interpreted thermodynamically by connecting them to the extra pressure $P_k$ and its conjugate volume $V_k$. So, we interpret the quantity 
\begin{equation}
P_k=-\frac{1}{8\pi k}
\end{equation}
as the pressure that arises due to the presence of the parameter $k\,$, in the similar way the pressure $P$ results from the cosmological constant $\Lambda$ and the quantity 
\begin{equation}
V_k= \left( \frac{\partial M}{\partial P_k} \right) _{S,P} = 4\pi k^2 e^{k/r} \left(1+ \frac{r_+^2}{l^2} \right)  
\end{equation}
 as its conjugate volume. The pressure $P_k$, termed as the tension of the self gravitating droplet of anisotropic fluid, works on the thermodynamic system by pushing it against gravitational collapse. Therefore, the two pressures $P$ and $P_k$ have opposite signs and affect the system in opposite manner.
The newly defined pair of conjugate variables ($P_k, V_k$) leads to the modification in the first law and Smarr relation 
\begin{eqnarray}
\mathrm{d} M &=& T \mathrm{d} S + V \mathrm{d} P + V_k \mathrm{d} P_k + \bar{\phi} dk,\,\\
M &=& 2 T S' - 2P V- 2 P_k V_k + \bar{\phi} k
.
\end{eqnarray}

 The Gibbs free energy bespeaks the global stability of the thermodynamical system and its global minimum indicates the preferred state, so we obtained the Gibbs free energy by using the relation $G_+=M_+-T_+S_+$ \cite{ak} as
\begin{eqnarray}
\tilde{G}_+&=&\frac{G_+}{l}=\frac{x_+}{4}\Big[e^{\frac{-\tilde{k}}{x_+}}\left(1-x_+^2+\frac{\tilde{k}}{x_+}(1+2x_+^2+3x_+^4)-\tilde{k}^2x_+(1+x_+^2)\right)\nonumber\\&&~~~~~~~~~~~~~~~~~~~-\frac{\tilde{k}^2}{x_+^2}Ei\left[\frac{\tilde{k}}{x_+}\right]\left(1+3x_+^2-\frac{\tilde{k}}{x}(1+x_+^2)\right)\Big],
\end{eqnarray} which reduces to the Gibb's free energy of Schwarzschild-AdS black holes \cite{Cai}, when $k==0$,
\begin{eqnarray}
\tilde{G}_+&=&\frac{x_+}{4}\left[1-x_+^2\right].
\end{eqnarray}
\begin{figure}[!h!h] 
\begin{tabular}{c c c c}
\includegraphics[width=0.5 \textwidth]{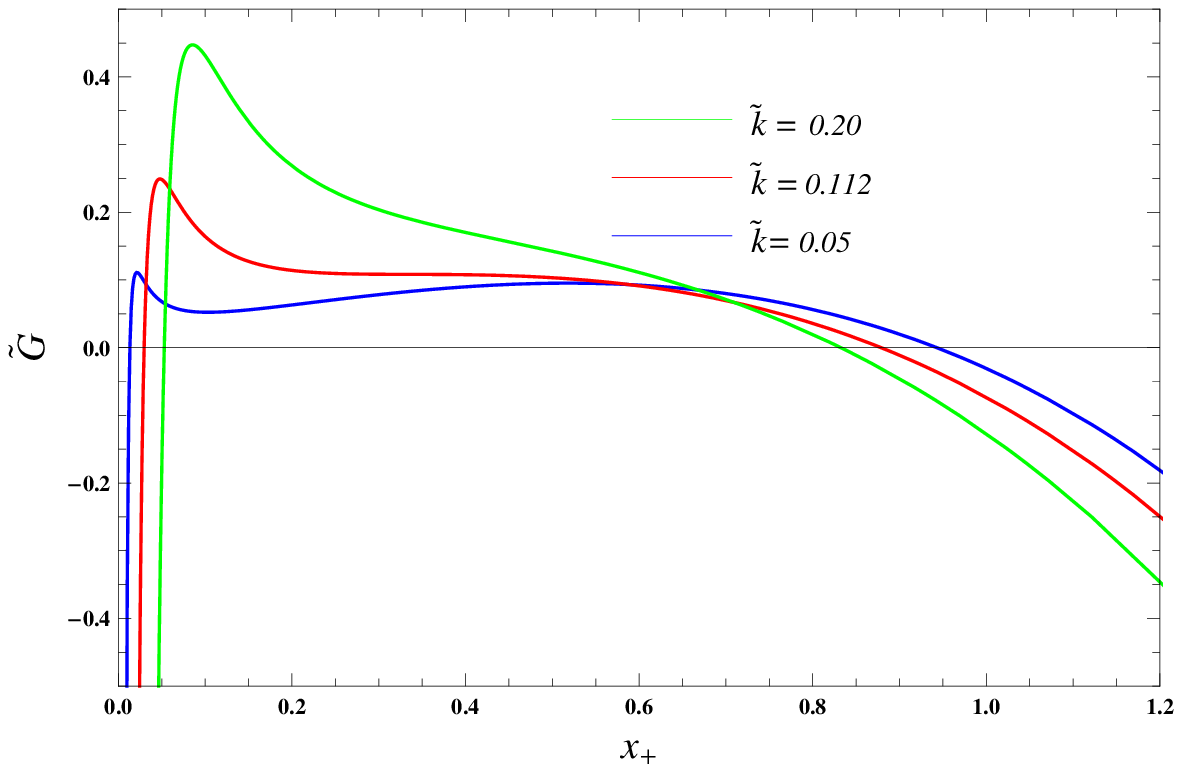}
\includegraphics[width=0.5 \textwidth]{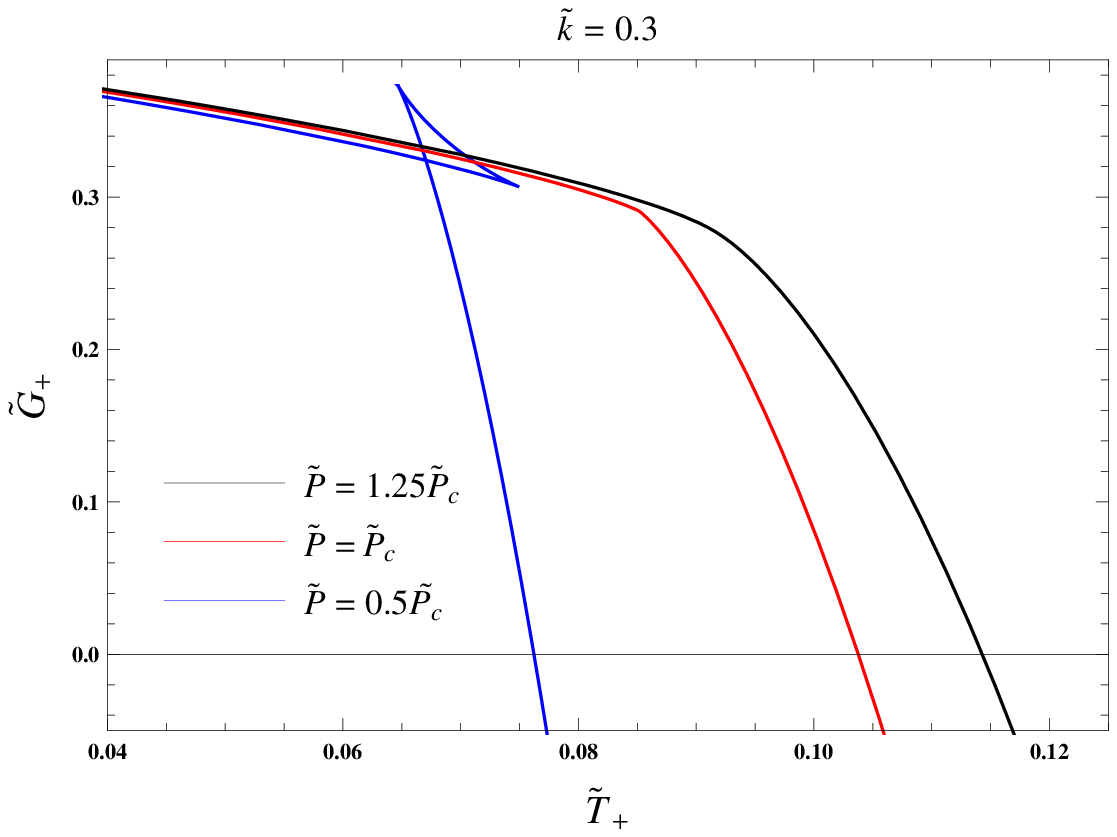}
\end{tabular}
\caption{ Gibb's free energy ($\tilde{G}_+$) \textit{vs} horizon $x_+$ (Left panel) and \textit{vs} temperature $\tilde{T_+}$ (Right panel).}
\label{Fig:gibbs}
\end{figure}
We have presented the behaviour of Gibbs free energy $\tilde{G}_+$ \textit{vs} horizon  $x_+$ and  temperature $\tilde{T_+}$ of nonsingular black hole in Fig.(\ref{Fig:gibbs}). We know that the black holes with negative Gibbs free energy are globally stable, hence, from Fig. (\ref{Fig:gibbs}) (Left panel), one can clearly say that the large black holes are globally stable, whereas, the black holes with small horizon radius are not stable globally. Fig. (\ref{Fig:gibbs}) (Right panel), shows the behaviour of Gibbs free energy $\tilde{G}_+$ \textit{vs} temperature $\tilde{T}_+$, in which one can see swallowtail-like structure when pressure is below the critical pressure $\tilde{P}_c$, which confirms the first order phase transition between small and large black holes \cite{dh}. Above the critical pressure $\tilde{P}_c$, we see no swallowtail-structure suggesting that no first order phase transition is happening. We also find that Gibbs free energy changes its sign which signifies the existence of a Hawking-Page phase transition between the thermal radiation and black hole \cite{HaP83}.

We also investigate the local thermodynamical stability of the nonsingular black hole through two specific heats ($C_{V}$ and $C_P$). It is well known that the sign of specific heats signifies the local stability of the black hole. If the specific heat is positive, then the black hole is locally stable, whereas, the negative specific heat signifies the local instability of the black hole. The divergence of the specific heat indicates that the phase transition exists. Two specific heats can be defined as, $C_V=T \left( \frac{\partial S}{\partial T} \right)_{V,k} \,,$ at constant volume and $C_P = T \left( \frac{\partial S}{\partial T} \right)_{P,k}$ at constant pressure. The specific heat ($C_V$) at constant volume turns out to be zero for the nonsingular black hole, whereas the expression of the specific heat at constant pressure ($C_P$) is given as
\begin{equation} \label{BAdSCp}
\tilde{C_P} =\frac{C_P}{l^2}=-2 \pi  x_+^2 \left[\frac{e^{\frac{\tilde{k}}{x_+}}\left(1+3x_+^2-\frac{\tilde{k}}{x_+}(x_+^2+1)\right)}{1-3x_+^2-2\frac{\tilde{k}}{x_+}}\right],
\end{equation}
 which goes over to the specific heat of Schwarzschild-AdS black holes in the absence of nonlinear electrodynamics
\begin{equation} \label{BAdSCp1}
\tilde{C_P} = -2 \pi  x_+^2 \left[\frac{1+3x_+^2}{1-3x_+^2}\right].
\end{equation}
\begin{figure}[!h] 
\centering
\includegraphics[width=0.6 \textwidth]{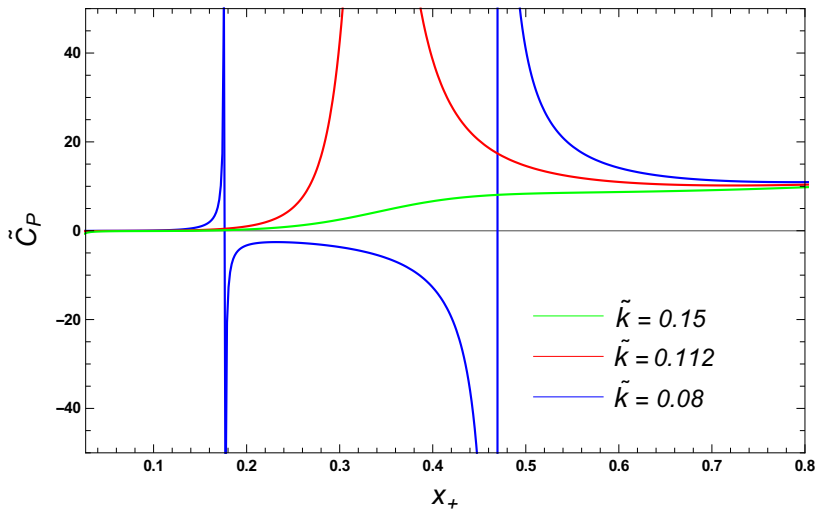}%
\caption{The specific heat ($\tilde{C_P}$) \textit{vs} horizon $x_+$.}
\label{Fig:hat}
\end{figure}
We plot $\tilde{C_P}$ in Fig.~\ref{Fig:hat}. As can be seen, when $\tilde{k}=0.8<\tilde{k}_c$ there are three branches divided by two asymptotes at $x_+\approx0.177$ and $x_+\approx0.471$, small black holes with $ x_+  \lesssim 0.177 \,$ and large black holes with $ x_+  \gtrsim 0.471 \,,$ for which $(\tilde{C_P}>0)$ signifying their local thermodynamical stability and the intermediate black holes with $0.177 \lesssim x_+ \lesssim 0.471$ having $\tilde{C_P}<0$ are thermodynamically unstable. It may be noteworthy that the specific heat suffers discontinuity at two points corresponding to maximum and minimum of temperature (cf. Fig. \ref{fig:Btemp}) confirming the second order phase transition. When $\tilde{k} = \tilde{k}_c=0.112$, we have two branches corresponding to small and large thermodynamically stable black holes which coexist at the inflexion point $x_+ \approx 0.338$ and for $\tilde{k}=0.15>\tilde{k}_c$, only thermodynamically stable black holes exist.

\begin{figure}[h!] 
\includegraphics[width=0.6 \textwidth]{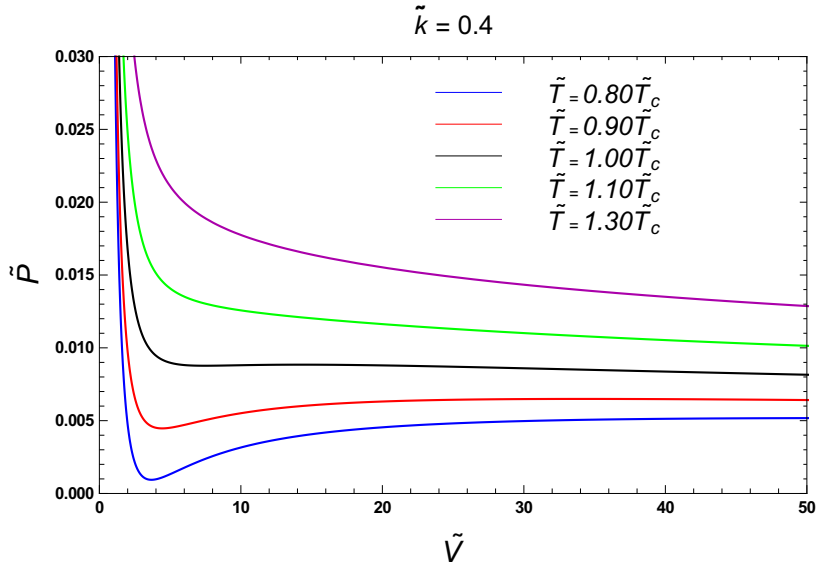}
\caption{The pressure $\tilde{P}$ \textit{vs} the volume $\tilde{V}$ for $\tilde{k}=0.4$. }
\label{Fig:pres}
\end{figure} 
Further, to check the $P-V$ criticality of the nonsingular black hole we obtain the equation of state $P=P(V,T)$ for the nonsingular black hole by combining the expressions of volume ($V$), temperature ($T$) and pressure ($P$) from Eq. \eqref{new_th}, as

\begin{equation} \label{Beos2}
P=\frac{3T}{3r_+-k}-\frac{3(r_+-k)}{8\pi r_+^2(3r_+-k)},
\end{equation} 
where $r_+$ is a function of thermodynamical volume $V$. The pressure of the system can be written in the form of dimensionless quantities as
\begin{equation}\label{pr}
\tilde{P}=Pl^2=\frac{3\tilde{T}}{3x_+-\tilde{k}}-\frac{3(x_+-\tilde{k})}{8\pi x_+^2(3x_+-\tilde{k})}.
\end{equation}
 We depict the isotherms of Eq. \eqref{Beos2} in the $P-V$ diagrams in Fig. \ref{Fig:pres}. One can easily notice that the system shows ideal gas behaviour, when the temperature is higher than the critical temperature $\tilde(T_c)$. The isotherm has an inflection point at $\tilde(T_c)$, which can be obtained by using

\begin{equation} \label{crtical_eq}
\left(\frac{\partial \tilde{P}}{\partial x_+}\right)_{\tilde{T}} = 0 = \left(\frac{\partial ^2 \tilde{P}}{\partial x_+^2}\right)_{\tilde{T}} \,.
\end{equation}
For $\tilde{T_+}<\tilde{T_c}$, the system undergoes the liquid-gas phase transition. The branch with high pressure belongs to small black holes, whereas, the branch with low pressure belongs to the large black holes. There exists an oscillating branch between the small and the large black holes, where the black undergoes the Van der Waals first order phase transition between the small and large black holes.
By using Eq. \eqref{pr} in Eq. \eqref{crtical_eq}, we can get the following critical temperature $\tilde{T_c}$ expressed in terms of the critical radius $x_c$ 
\begin{equation}\label{t_c}
\tilde{T}_c=T_c l=\frac{3 x_c^2-5 \tilde{k} x_c+\tilde{k}^2}{6 \pi  x_c^3},
\end{equation} 
and further we can obtain the expression for critical pressure $\tilde{P}_c$, in terms of the critical radius $x_c$
\begin{equation}\label{p_c}
\tilde{P}_c=\frac{x_c-2 \tilde{k}}{8 \pi  x_c^3}
\end{equation}
Now by inserting the value of critical radius $x_c=3\tilde{k}$, in Eqs. \eqref{t_c}, \eqref{p_c} and in the expression of the volume from Eq. \eqref{new_th}, we get the following critical values of temperature, pressure and volume
\begin{equation}\label{cr}
\tilde{P}_c \simeq \frac{1}{216 \pi  \tilde{k}^2}\,, \ \ \ \ \tilde{V}_c \simeq 36 \sqrt[3]{e} \pi  \tilde{k}^3\,, \ \ \ \ \tilde{T}_c \simeq \frac{13}{162 \pi  \tilde{k}} \,,
\end{equation}

The point at which pressure, volume and temperature have the critical values as given in Eq. \eqref{cr}, known as the critical point, is the point at which we cannot distinguish small and large black holes (small and large black holes coexist). The universal constant for nonsingular black hole is obtained as
 \begin{equation}
 \varepsilon = \frac{\tilde{P_c}\tilde{ V_c}^{1/3}}{\tilde{T_c}} \approx 0.31,
 \end{equation}
 which is slightly smaller than the value $3/8$ of the Van der Waals gas. 
 \section{Critical Exponents}
 Further, we evaluate the critical exponents $\alpha, \beta, \gamma$ and $\delta$, which describe the behaviour of the thermodynamic quantities near the critical point, through the following relations
 \begin{eqnarray} \label{crit_a}
C_V &=& T \frac{\partial S}{\partial T} \Big |_V \propto |t|^{-\alpha} \,, \\ \label{crit_b}
\eta &=& V_l - V_s \propto |t|^{\beta} \,, \\ \label{crit_c}
\kappa_T &=& - \frac{1}{V} \frac{\partial V}{\partial P} \Big |_T \propto |t|^{-\gamma} \,, \\ \label{crit_d}
|P-P_c|_{T=T_c}  & \propto &  |V-V_c|^{\delta} \,,
\end{eqnarray}
with $t=(T-T_c)/T $. Here, $C_V, \eta, \kappa_T $ and $|P-P_c|_{T=T_c}$ are respectively, the specific heat at constant volume, the order parameter, the isothermal compressibility and the critical isotherm with $V_l$ and $V_s$ are the volume of large and small black holes, respectively. The specific heat at constant volume for nonsingular black holes is equal to zero, leading to $\alpha=0$.

We define the following dimensionless quantities to study the behaviour of the physical quantities near the critical point
\begin{eqnarray}\label{dimqun}
p=\frac{\tilde{P}}{\tilde{P}_c},~~~~~~~~~1+\epsilon=\frac{x_+}{x_c},~~~~~~~~~~~~~1+\omega=\frac{\tilde{V}}{\tilde{V}_c},~~~~~~~~~~~~1+t=\frac{\tilde{T}}{\tilde{T}_c},
\end{eqnarray} with $|\epsilon|, |\omega|, |t| \ll 1.$ We can obtain the relation between between $\omega$ and $\epsilon$, by inserting the above expression of $x_+$ in the expression of volume from Eq. \eqref{new_th} as
\begin{equation}\label{46}
\omega=4\left(1-\frac{\tilde{k}}{2x_c}\right)\epsilon.
\end{equation}
We can write the equation of state in terms of dimensionless quantities defined in Eq. (\ref{dimqun}) as
\begin{equation}\label{eqnst}
p=1+At-Bt\epsilon -C\epsilon^3+\mathcal{O}(t\epsilon^2,\epsilon^4),
\end{equation} with
\begin{eqnarray}
A&=&\frac{3}{2(3x_c-\tilde{k})}\frac{\tilde{T}_c}{\tilde{P}_c},~~~~~~~~B=\frac{9x_c}{2(3x_c-\tilde{k})^2}\frac{\tilde{T}_c}{\tilde{P}_c},\nonumber\\
C&=&\frac{1}{8 \pi x_c^2(3x_c-\tilde{k})^4\tilde{P_c}}\Big[-162x_c^4+27(21-6\tilde{k})x_c^3-18\tilde{k}(18-k)x_c^2+3\tilde{k}^2(45+\tilde{k})x_c-12\tilde{k}^3\Big].\nonumber\\
\end{eqnarray}
Further, to calculate the remaining critical exponential, we take fixed $t<0$. We know the pressure of the system remains constant during the phase transition (small to large black holes), so we can write in the form of following equation
\begin{equation}\label{49}
1+At-Bt\epsilon_{s} -C\epsilon_{s}^3=1+At-Bt\epsilon_{l} -C\epsilon_{l}^3,
\end{equation} where $\epsilon_s$ and $\epsilon_l$ respectively correspond to the radii of small and large black holes. To obtain the value of critical exponential $\beta$, we use the well known Maxwell's area law
\begin{equation}\label{50}
\int_{\omega_s}^{\omega_l}\omega d\tilde{P}=0.
\end{equation}
 By differentiating Eq. (\ref{eqnst}), we get
 \begin{equation}\label{51}
 d\tilde{P}=-\tilde{P}_c(Bt+3C\epsilon^2)d\epsilon.
 \end{equation}
 Now, using Eq. (\ref{51}) and Eq. (\ref{46}) in Eq. (\ref{50}), one can obtain
 \begin{equation}\label{52}
 \int_{\epsilon_s}^{\epsilon_l}\omega d\tilde{P}=-4\tilde{P}_c(1-\frac{\tilde{k}}{2x_c^2})\int_{\epsilon_s}^{\epsilon_l}\epsilon(Bt+3C\epsilon^2)d\epsilon.
\end{equation}  
By solving Eq. (\ref{49}) and Eq. (\ref{52}), one can get the following unique solution
\begin{equation}
\epsilon_l=-\epsilon_s=\sqrt{\frac{-Bt}{C}}.
\end{equation}
Thus Eq. (\ref{crit_b}), can be written as
\begin{equation}
\eta=\tilde{V}_l-\tilde{V}_s=\tilde{V}_c(\omega_l-\omega_s)\propto \sqrt{-t},
\end{equation} which leads to $\beta=1/2$. To obtain the critical exponent $\gamma$, we need to calculate the thermal compressibility using Eq. (\ref{crit_c}),
\begin{equation}
\kappa_T = - \frac{1}{\tilde{V}} \frac{\partial \tilde{V}}{\partial\tilde{P}}|_{\tilde{T}}=\frac{4}{\tilde{P}_c(1+\omega)}(1-\frac{\tilde{k}}{2x_c^2})(\frac{\partial \epsilon}{\partial p}|_t)\propto\frac{1}{Bt},
\end{equation} 
signifyng $\gamma=1$. To obtain the value of $\delta$, we compute $|P-P_{c}|$ at $T=T_c$ or $t=0$, as
\begin{equation}
|\tilde{P}-\tilde{P}_c||_{\tilde{T}_c}\propto|\epsilon|^3\propto|\tilde{V}-\tilde{V}_c|^3.
\end{equation} Hence, we get $\delta=3$. Here one can notice that the values of the critical exponents of our nonsingular-AdS black holes are the same values of the critical exponents in the Van der Waals fluid \cite{dr1,ac}.
\section{Final remarks}
In this paper, we have obtained exact nonsingular black holes in AdS spacetime which encompasses the Schwarzschild-AdS as a special case in the absence of nonlinear electrodynamics ($ k = 0 $) and asymptotically ($x\gg x$)  mimics the well known Reissner-Nordstrom-AdS black hole.  Thus, we have constructed an exact spherically symmetric, static,  regular black hole metrics in the context of general relativity minimally coupled to nonlinear electrodynamics theory.  The new nonsingular-AdS black hole metric is characterized by horizons which could be at least two describing a variety of self-gravitating objects, including an extremal black hole with degenerate horizons and non-extremal black holes with two distinct horizons.   We have shown that the exponential correction term arising due to nonlinear electrodynamics can cure the singularity problem and analyze in the detailed thermodynamical properties of the obtained nonsingular-AdS black hole metric were investigated.  We considered that the cosmological constant plays the role of the positive pressure of the system. After having calculated the desired thermodynamic variables, we discussed the thermodynamical quantities like the Hawking temperature, entropy, specific heat at constant pressure and Gibb's free energy in extended phase space.   The resultant forms of the specific heat and the Gibbs free energy indicate a first-order phase transition between small/large black hole.  The $P$ vs $r_+$ diagram shows that below the critical pressure $P_c$,  the nonsingular-AdS black hole undergoes phase transition between small/large black hole reminiscent of the liquid/gas
coexistence. Finally, we discussed the $P-V$ criticality for our nonsingular-AdS black hole, and by analysing the behaviour of various thermodynamical quantities near the critical point, we found the values of the critical exponents and showed that these values are the same as those for the Van der Waals gas.

\subsection*{Acknowledgments}
S.G.G. and S.D.M.  would like to thank DST INDO-SA bilateral project DST/INT/South Africa/P-06/2016, S.G.G. also thank SERB-DST for the ASEAN project IMRC/AISTDF/CRD/2018/000042.  S.D.M. acknowledges
that this work is based upon research supported by the South African
Research Chair Initiative of the Department of Science and
Technology and the National Research Foundation.
S.G.G. would like to also thank Rahul Kumar for fruitful discussions and IUCAA, Pune for the hospitality while this work was being done.

\end{document}